\documentclass{article}
\usepackage{arxiv}

\usepackage{filecontents}

\usepackage{tikz}
\usetikzlibrary{arrows,positioning,shadows}
\tikzset{
    >=stealth',
    punkt/.style={
           rectangle,
           rounded corners,
           draw=black, very thick,
           text width=6.5em,
           minimum height=2em,
           text centered},
    pil/.style={
           ->,
           thick,
           shorten <=2pt,
           shorten >=2pt,}
}
\usepackage{pgf-umlsd}

\usetikzlibrary{fit}
\tikzset{
  comp/.style = {
    minimum width  = 2cm,
    minimum height = 1cm,
    text width     = 2cm,
    inner sep      = 0pt,
    text           = white,
    align          = center,
    font           = \normalsize,
    transform shape,
    thick
  },
  monitor/.style = {draw = none, xscale = 18/16, yscale = 11/9},
  display/.style = {shading = axis, left color = black!60, right color = black},
  ut/.style      = {fill = gray}
}
\tikzset{
  computer/.pic = {
    \node(-m) [comp, pic actions, monitor]
      {\phantom{\parbox{\linewidth}{\tikzpictext}}};
    \node[comp, pic actions, display] {\tikzpictext};
    \begin{scope}[x = (-m.east), y = (-m.north)]
      \path[pic actions, draw = none]
        ([yshift=2\pgflinewidth]-0.1,-1) -- (-0.1,-1.3) -- (-1,-1.3) --
        (-1,-2.4) -- (1,-2.4) -- (1,-1.3) -- (0.1,-1.3) --
        ([yshift=2\pgflinewidth]0.1,-1);
      \path[ut]
        (-1,-2.4) rectangle (1,-1.3)
        (-0.9,-1.4) -- (-0.7,-2.3) -- (0.7,-2.3) -- (0.9,-1.4) -- cycle;
      \path[pic actions, fill = none]
        (-1,1) -- (-1,-1) -- (-0.1,-1) -- (-0.1,-1.3) -- (-1,-1.3) --
        (-1,-2.4) coordinate(sw)coordinate[pos=0.5] (-b west) --
        (1,-2.4) -- (1,-1.3) coordinate[pos=0.5] (-b east) --
        (0.1,-1.3) -- (0.1,-1) -- (1,-1) -- (1,1) -- cycle;
      \node(-c) [fit = (sw)(-m.north east), inner sep = 0pt] {};
    \end{scope}
  }
}

\usepackage{pgfplots}
\usetikzlibrary{shapes}
\usepackage{caption}
\usepackage{subcaption}

\usepackage[nolist,nohyperlinks]{acronym}

\usepackage{setspace}

\usepackage{amssymb}
\usepackage{pifont}
\newcommand{\cmark}{\ding{51}}%

\usepackage{tabulary}
\newcolumntype{K}[1]{>{\arraybackslash}p{#1}}
\usepackage{courier}

\usepackage{graphicx}
\usepackage{float}

\usepackage{url}

\newcommand\copyrighttext{%
  \footnotesize \textcopyright 2019 IEEE. 
  Personal use of this material is permitted.
  Permission from IEEE must be obtained for all other uses,
  in any current or future media,
  including reprinting/republishing this material for advertising or promotional purposes,
  creating new collective works, for resale or redistribution to servers or lists,
  or reuse of any copyrighted component of this work in other works.
  DOI: 10.23919/AE.2017.8053600 -- \url{http://dx.doi.org/10.23919/AE.2017.8053600}
}
\newcommand\copyrightnotice{%
\begin{tikzpicture}[remember picture,overlay]
\node[anchor=south,yshift=10pt] at (current page.south) {\fbox{\parbox{\dimexpr\textwidth-\fboxsep-\fboxrule\relax}{\copyrighttext}}};
\end{tikzpicture}%
}

\title{PropFuzz - An IT-Security Fuzzing Framework for Proprietary ICS Protocols}

\author{
 Matthias Niedermaier \\
 Matthias.Niedermaier@hs-augsburg.de \\
 Hochschule Augsburg
 \And
 Florian Fischer \\
 Florian.Fischer@hs-augsburg.de \\
 Hochschule Augsburg
 \And
 Alexander von Bodisco\\
 Alexander.vonBodisco@hs-augsburg.de \\
 Hochschule Augsburg
}

\begin{document}

\maketitle

\begin{abstract}
Programmable Logic Controllers are used for smart homes, in production processes or to control critical infrastructures.
Modern industrial devices in the control level are often communicating over proprietary protocols on top of \acs{TCP}/\acs{IP} with each other and SCADA systems.
The networks in which the controllers operate are usually considered as trustworthy and thereby they are not properly secured.
Due to the growing connectivity caused by the \ac{IoT} and Industry 4.0 the security risks are rising.
Therefore, the demand of security assessment tools for industrial networks is high.
In this paper, we introduce a new fuzzing framework called PropFuzz, which is capable to fuzz proprietary industrial control system protocols and monitor the behavior of the controller.
Furthermore, we present first results of a security assessment with our framework.
\end{abstract}

\copyrightnotice
\acresetall

\section{Introduction}
\label{sec:introduction}
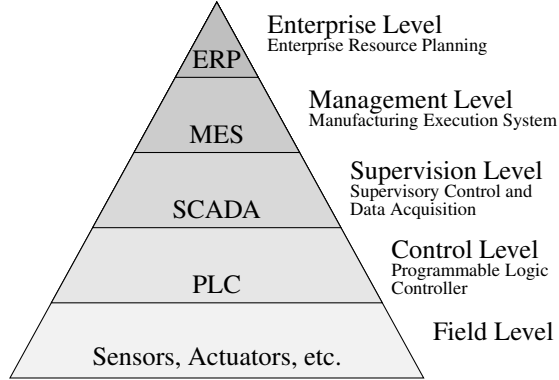
\begin{figure}[htb]
  \centering
\begin{tikzpicture}
\coordinate (A) at (-3.3,-1) {};
\coordinate (B) at (3.3,-1) {};
\coordinate (C) at (0,5) {};
\coordinate (D) at (5,0) {};
\coordinate (E) at (3.3,-1) {};
\foreach \A/\col [count=\i,evaluate=\i as \j using 5*\i] in {{Sensors, Actuators, etc.}, PLC/PAC, SCADA, MES, ERP}
\draw[fill=black!\j] (C)--([shift={(-.55*\i,1*\i)}]B)--node[above,align=center] {\A}([shift={(.55*\i,1*\i)}]A)--cycle;
\foreach \A/\col [count=\i,evaluate=\i as \j using 5*\i] in {Field Level \\ , Control Level \\[-0.15cm]  \scriptsize{Programmable Logic} \\[-0.15cm] \scriptsize{Controller}, Supervision Level \\[-0.15cm]   \scriptsize{Supervisory Control and} \\[-0.15cm]  \scriptsize{Data Acquisition}, Management Level \\[-0.15cm]   \scriptsize{\aclu{MES}} \\[-0.15cm], Enterprise Level \\[-0.15cm]  \scriptsize{\aclu{ERP}} \\[-0.15cm]}
\draw ([shift={(-.55*\i,1*\i)}]E) node[above right,align=left] {\A}([shift={(.55*\i,1*\i)}]D);
\end{tikzpicture}
\caption{Common Topology for Industrial Networks}
\label{fig_topology}
\end{figure}

\acp{PLC} are the basic components used in a wide variety of \acp{ICS} for example to regulate production processes or to control critical infrastructures.
Historically, these devices operated in a separated network, with no connection to the Internet or office areas.
In the past decade, this separation changed due to the demand of highly connected systems in the \ac{IoT} and the fourth industrial revolution.
A typical company network with control systems consists of several hierarchical layers as illustrated in Figure \ref{fig_topology}.
Nowadays, the communication on the higher levels (\acs{ERP}, \acs{MES}, \acs{SCADA} and \acs{PLC}) is mostly based on the \ac{IP} protocol.

Most \ac{PLC}s offer the possibility to configure and program them via a proprietary \acs{TCP}/\acs{IP} connection.
This simplification allows remote access to these devices, if there is no additional hardware restricting the communication.
Thus, it is often possible for attackers to interact directly with the \ac{PLC}.
Therefore, it is necessary to analyze the communication between the control system and the \ac{IDE} with the aim of fuzzing proprietary industrial protocols to find security issues.
The main problem concerning fuzzing these protocols is defining the data structure to be fuzzed.

In our fuzzing framework, the inputs are identified by using statistical computation to analyze the structure of proprietary industrial protocols.
Popular fuzzing frameworks are introduced in Section \ref{sec:relatedwork}.
Section \ref{sec:methodology} describes the methodology of fuzzing a proprietary \ac{PLC} communication.
Section \ref{sec:framework} explains our framework architecture.
In Section \ref{sec:assesment} and \ref{sec:recommendations} first result, with possible attacks and recommendations are presented.
Finally, an outlook and conclusion in Section \ref{sec:outlook} is given.

\section{Related Work and Motivation}
\label{sec:relatedwork}
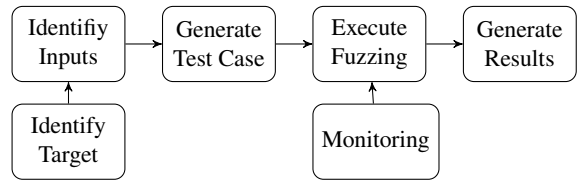
\begin{figure}[H]
  \centering
 \begin{tikzpicture}[%
    auto,
    block/.style={
      rectangle,
      draw=black,
      align=center,
      rounded corners
    }
  ]
  \node[block, align=center, minimum width=1.5cm, minimum height=1cm, anchor=south west] at (0, -1.3)   (a) {\footnotesize Identify \\ \footnotesize Target};
  \node[block, align=center, minimum width=1.5cm, minimum height=1cm, anchor=south west] at (0, 0)   (b) {\footnotesize Identifiy \\ \footnotesize Inputs};
  \node[block, align=center, minimum width=1.5cm, minimum height=1cm, anchor=south west] at (2, 0)   (c) {\footnotesize Generate \\ \footnotesize Test Case};
  \node[block, align=center, minimum width=1.5cm, minimum height=1cm, anchor=south west] at (4, 0)   (d) {\footnotesize Execute \\ \footnotesize Fuzzing};
  \node[block, align=center, minimum width=1.5cm, minimum height=1cm, anchor=south west] at (6, 0)   (e) {\footnotesize Generate \\ \footnotesize Results};
  \node[block, align=center, minimum width=1.5cm, minimum height=1cm, anchor=south west] at (4, -1.3)   (f) {\footnotesize Monitoring};

  \draw [->] (a) -- (b);
  \draw [->] (b) -- (c);
  \draw [->] (c) -- (d);
  \draw [->] (d) -- (e);
  \draw [->] (f) -- (d);
  \end{tikzpicture}
  \caption{Fuzzing Test Process}
  \label{fig_fuzzingtest}
\end{figure}

Barton Miller, discovered a program crash caused by noise as a result of a lightning strike on his network connection during a thunderstorm \cite{Miller:1990:ESR:96267.96279}.
This bug was triggered by a random input which is called fuzz-testing or fuzzing in the literature.
Fuzzing could only trigger bugs, if the input is not rejected by a validation function of the \ac{DuT}.
A full automated fuzzing framework for \ac{ICS}s include the process steps illustrated in Figure \ref{fig_fuzzingtest} \cite{7560424}.

There are two elementary categories of fuzzers, based on how they create input for fuzzing.
Generation-based fuzzers create input from scratch and thus require some knowledge of the protocol with corresponding data fields.
With mutation fuzzers, samples of valid input are used to produce malformed input.
A simple mutation fuzzer can modify a valid input sample and send it to the \ac{DuT}.
\begin{itemize}
\item \textbf{Generation-Based Fuzzing} applies with generation rules to fuzz input.
BooFuzz \cite{Boofuzz}, a fork and successor of the Sulley \cite{sulley} fuzzing framework, and Peach \cite{Peachfuzz} are block-based fuzzers.
These kinds of fuzzers require a deep knowledge of the protocol structure and test case definition to generate inputs.
Recent generation-based fuzzer like VUzzer \cite{rawat2017vuzzer} are able to automatically generate input test cases for basic communications.
\item \textbf{Mutation Fuzzing} uses valid inputs and modifies them to create fuzzing input.
Most of the frameworks analyze previously captured traffic, although there are fuzzers which allow live capturing.
Radasma \cite{Radasma} is an input generation tool for basic protocols to identify field boundaries.
LZFuzz framework \cite{Shapiro2011} is an online fuzzer, which intercepts traffic directly, analyzes it with the Lempel–Ziv compression algorithm and sends the manipulated packages to the device.
\end{itemize}

For fuzz-testing \ac{ICS}s, these fuzzers and frameworks do not fulfill our requirements in automated input generation for proprietary protocols and electrical monitoring \cite{Shapiro2011}.
The documentation between the \ac{IDE} and the \ac{PLC} is mostly not public available and reverse engineering is highly time-consuming.
Thus, it is necessary to have an automatic fuzzing framework for this kind of communication.
More over in case of \ac{ICS} it is necessary to monitor any behavior change of the device, where network monitoring is not enough.

\section{Methodology}
\label{sec:methodology}
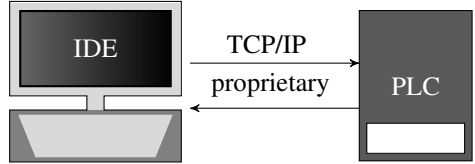
\begin{figure}[H]
  \centering
\begin{tikzpicture}
  \pic(comp0) [
    draw,
    fill = gray!30,
    pic text = {\ac{IDE}}
  ]
  {computer};

  \node[draw, align=center, minimum width=1.5cm, minimum height=2cm, anchor=south west, fill=black!65!white] at (3.5, -1.5)   (a) {\textcolor{white}{\ac{PLC}}};
  \node[draw, align=center, minimum width=1.3cm, minimum height=0.4cm, anchor=south west, fill=white] at (3.6, -1.4)   (b) {};

  \draw [->] ([xshift=-3cm, yshift=0.3cm]a.center) -- ([xshift=-0.75cm, yshift=0.3cm]a.center);
  \draw [->] ([xshift=-0.75cm, yshift=-0.3cm]a.center) -- ([xshift=-3cm, yshift=-0.3cm]a.center);

  \node[text width=2cm, align=right, anchor=north] at (1.8,0.3) {TCP/IP};
  \node[text width=2cm, align=right, anchor=north] at (2.1,-0.2) {proprietary};
\end{tikzpicture}
\caption{Communication between \ac{IDE} and \ac{PLC}}
\label{fig_network}
\end{figure}

Most of the modern \ac{PLC}s are programmed with an \ac{IDE} over \acs{TCP}/\acs{IP}.
This communication is often open and not filtered.
Figure \ref{fig_network} illustrates the minimal setup to interact with the \ac{ICS} and the \ac{IDE}.

The majority of these protocols are proprietary and have no public available documentation.
Our framework allows a direct investigation of this communication.
To start the data transfer between an \ac{IDE} and a \ac{PLC} a \acs{TCP}/\acs{IP} handshake is done.
After that, there is often an additional proprietary handshake with a kind of challenge.
Followed by this, the command and data transfer could be started.
For a permanent connection, it could be essential to send keep-alive messages between the \ac{IDE} and the \ac{PLC}.
This is not required for single command interaction.
To make fuzzing feasible, it is necessary to perform the proprietary handshake and determine the protocol field, which should be fuzzed.

\section{Framework Architecture}
\label{sec:framework}
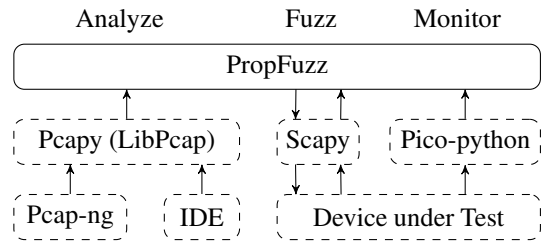
\begin{figure}[H]
  \centering
\begin{tikzpicture}[node distance=1cm,
    auto,
    block/.style={
      rectangle,
      draw=black,
      align=center,
      rounded corners,
      dashed
    }
  ]
    \node[block, draw, align=center, minimum width=1.5cm, minimum height=0.6cm, anchor=south west] at (0, 0)   (a) {Pcap-ng};
    \node[block, draw, align=center, minimum width=1cm, minimum height=0.6cm, anchor=south west] at (2, 0)   (b) {IDE};
    \node[block, draw, align=center, minimum width=3cm, minimum height=0.6cm, anchor=south west] at (0, 1)   (c) {Pcapy (LibPcap)};
    \node[block, solid, draw, fill=white, align=center, minimum width=7.0cm, minimum height=0.6cm, anchor=south west] at (0, 2)   (d) {PropFuzz};
    \node[block, draw, align=center, minimum width=1cm, minimum height=0.6cm, anchor=south west] at (3.5, 1)   (e) {Scapy};
    \node[block, draw, align=center, minimum width=2cm, minimum height=0.6cm, anchor=south west] at (5, 1)   (f) {Pico-python};
    \node[block, draw, align=center, minimum width=3.5cm, minimum height=0.6cm, anchor=south west] at (3.5, 0)   (g) {Device under Test};

    \node[text width=2cm, align=left, anchor=north west] at (0.7,3.2) {Analyze};
    \node[text width=2cm, align=right, anchor=north] at (3.3,3.2) {Fuzz};
    \node[text width=2cm, align=right, anchor=north east] at (6.6,3.2) {Monitor};

	\draw [->] (a) -- ([xshift=-0.75cm, yshift=-0.3cm]c.center);
	\draw [->] (b) -- ([xshift=1.0cm, yshift=-0.3cm]c.center);
	\draw [->] (c) -- ([xshift=-2cm, yshift=-0.3cm]d.center);
	\draw [->] ([xshift=0.3cm, yshift=0.3cm]e.center) -- ([xshift=0.85cm, yshift=-0.3cm]d.center);
	\draw [->] ([xshift=0.25cm, yshift=-0.3cm]d.center) -- ([xshift=-0.3cm, yshift=0.3cm]e.center);
	\draw [->] ([xshift=-0.3cm, yshift=-0.3cm]e.center) -- ([xshift=-1.5cm, yshift=0.3cm]g.center);
	\draw [->] ([xshift=-0.9cm, yshift=0.3cm]g.center) -- ([xshift=0.3cm, yshift=-0.3cm]e.center);
	\draw [->] ([xshift=0.75cm, yshift=0.3cm]g.center) -- (f);
    \draw [->] (f) -- ([xshift=2.5cm, yshift=-0.3cm]d.center);
\end{tikzpicture}
\caption{Data-flow within PropFuzz Framework}
\label{fig_highlvl}
\end{figure}

At a high-level view, illustrated in Figure \ref{fig_highlvl}, PropFuzz is separated in three parts to fulfill the requirements of a full integrated fuzzing framework \cite{7577276}.
The \textbf{analyze} part splits the protocol and filtrates the information necessary to \textbf{fuzz} the \ac{DuT}.
In addition to the network response monitoring the \textbf{monitor} part observes the \ac{PLC} electrically.

\begin{figure}[H]
  \centering
\begin{tikzpicture}[node distance=1cm,
    auto,
    block/.style={
      rectangle,
      draw=black,
      align=center,
      rounded corners
    }
  ]
    \node[block, align=center, minimum width=3.1cm, minimum height=2.8cm, anchor=south west] at (0, 0)   (a){};
    \node (b) [cylinder, shape border rotate=90, draw,minimum height=1.5cm,minimum width=1.4cm, shape aspect=0.2, anchor=south west] at (0.2, 1.1) {Classes};
    \node (c) [cylinder, shape border rotate=90, draw,minimum height=1.5cm,minimum width=1.4cm, shape aspect=0.2, anchor=south west] at (1.7, 1.1) {Config};
    \node[block, align=center, minimum width=2.9cm, minimum height=0.7cm, anchor=south west] at (0.1, 0.1)   (d) {PropFuzz};

    \node[block, align=center, minimum width=2cm, minimum height=0.6cm, anchor=south west] at (4.5, 2.2)   (e) {Unpack};
    \node[block, align=center, minimum width=2cm, minimum height=0.6cm, anchor=south west] at (4.5, 1.5)   (f) {Analyze};
    \node[block, align=center, minimum width=2cm, minimum height=0.6cm, anchor=south west] at (4.5, 0.8)   (g) {SendFuzz};
    \node[block, align=center, minimum width=2cm, minimum height=0.6cm, anchor=south west] at (4.5, 0.1)   (h) {Monitor};

    \path[->]    ([yshift=0.15cm]a.east)  edge [bend left] (e.west);
    \path[->]    ([yshift=0.05cm]a.east)  edge [bend left] (f.west);
    \path[->]    ([yshift=-0.05cm]a.east)  edge [bend right] (g.west);
    \path[->]    ([yshift=-0.15cm]a.east)  edge [bend right] (h.west);
\end{tikzpicture}
\caption{PropFuzz Modularity}
\label{fig_implement}
\end{figure}
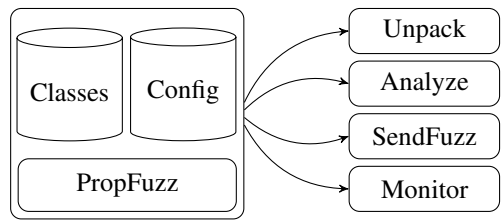

A detailed view of the PropFuzz structure is shown in Figure \ref{fig_implement}, which illustrates the python modules, classes and configuration files within PropFuzz.

The implemented python modules are modular and could be extended or used within other frameworks.

\subsection{Unpack}
The PropFuzz implementation provides two possibilities for package analysis.
First a live capturing of the communication between \ac{DuT} and \ac{IDE} is possible.
The second option is to read in existing pcap files.
For live capturing \ac{ARP} spoofing is needful, which could be prevented by some network devices.
Once data acquisition is done the Unpack module splits up the information of the captured packages and stores them in objects.
The module uses pcapy \cite{pcapy}, a python implementation for LibPcap \cite{libpcap}, for further examination of the collected packages.

\subsection{Analyze}
Inside the Analyze process the messages of the proprietary protocol are interpreted.
At the beginning a statistical analysis with the Ratcliff/Obershelp \cite{ratcliff1998ratcliff} pattern recognition algorithm of the created package objects is done.
With these similarities, the proprietary handshake between the \ac{IDE} and the \ac{PLC} can be determined.
After the detection of a handshake in the captured communication commands between the \ac{IDE} and the \ac{PLC} must be identified.
Commands can be identified by comparing different captures containing a full match of the same command.

\subsection{SendFuzz}
The SendFuzz module is responsible for sending and receiving packages inside the PropFuzz implementation.
The gained information from the Analysis module is used to mimic the protocol handshake by sending sniffed messages to the \ac{DuT}.
For constructing these messages the python module scapy \cite{Scapy} is used.
After a successful established protocol handshake, further packages containing protocol-specific commands are sent to the \ac{DuT}.

\subsection{Monitor}
Most fuzzing frameworks only observe the network connection during the test.
Special about fuzzing \ac{ICS}s is the monitoring of the process control \cite{7778820}.
To detect the effect of fuzz-testing with our framework, an output channel is monitored by an oscilloscope.
Therefore, a square wave signal is generated on a certain output of the \ac{DuT}, created with an alternating write to the output within the execution cycle of the \ac{PLC}.
Control commands sent to the \ac{PLC} which e.g. stop or restart the device make an impact on the periodic signal of the output.
This variation can then be monitored with the oscilloscope.

\begin{figure}[H]
  \centering
\begin{tikzpicture}
\begin{axis}[
width=8cm,
height=3cm,
x axis line style={-stealth},
y axis line style={-stealth},
xtick={1,2,3,4,5,6,7,8},
xticklabels={1,2,3,4,5,6,7,8},
ymax = 30,xmax=7.5,
axis lines*=center,
ytick={12,24},
xlabel={Time $\rightarrow$},
ylabel={Voltage (V)},
xlabel near ticks,
ylabel near ticks]
\addplot+[dotted, thick,mark=none,const plot, color=black!60!white]
coordinates
{(0,0) (0,24) (1,0) (2,24) (2,0) (2,24) (3,0) (4,24) (5.5,0) (6.5,24) (7.5,24)};
\addplot+[thick,mark=none,const plot, color=black]
coordinates
{(0,0) (0,24) (1,0) (2,24) (2,0) (2,24) (3,0) (4,24) (5,24)};
\end{axis}
\node[text width=1cm, align=right, anchor=north] at (2.5,1.7) (a) {\textcolor{black}{Expected}};
\node[text width=1cm, align=right, anchor=north] at (5.2,1.7) (a) {\textcolor{black!60!white}{Delayed}};
\end{tikzpicture}
\caption{Monitor \ac{PLC} Output with Oscilloscope}
\label{fig_freq}
\end{figure}
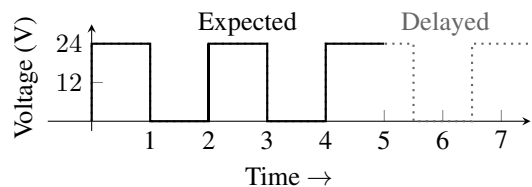

Figure \ref{fig_freq} illustrates the expected square wave (solid black) and an unexpected delayed output (\textcolor{black!60!white}{dotted grey}).
This behavior can occur if a command causes high load on the controller.
With our fuzzing framework, it is possible to trigger and monitor such changes.
If the delay exceeds the allowed output jitter the process control is not feasible. This could result in unexpected behavior.
For our framework, Pico-python \cite{Picopython} is used, to interact with the PicoScope 2208, which is a scriptable \ac{USB} oscilloscope.

\section{Security Assessment Evaluation}
\label{sec:assesment}
For testing PropFuzz two different \ac{PLC}s, the "ILC 171 ETH" and the "ILC 150 ETH" from Phoenix Contacts, are used.
According to the datasheet the devices support the protocols shown in  Table \ref{tab_ports}.
In this assessment the \ac{IDE} "AUTOMATIONWORX Software Suite v1.83" from Phoenix Contacts is used.

\begin{table}[H]
\centering
\caption{Phoenix Contacts Test Equipment}
\label{tab_ports}
\begin{tabular}{l c c}
\hline \hline
\textbf{DuT}     & \textbf{ILC 171} & \textbf{ILC 150} \\
\hline
Man. number      & 2700975                  & 2985330              \\
Profinet         & \cmark                   &                      \\
Modbus           & \cmark                   &                      \\
Proprietary      & \cmark                   & \cmark               \\
FTP              & \cmark                   & \cmark               \\
HTTP             & \cmark                   & \cmark               \\
HTTPS            & \cmark                   &                      \\
SNTP             & \cmark                   & \cmark               \\
SNMP             & \cmark                   & \cmark               \\
SMTP             & \cmark                   & \cmark               \\
SQL              & \cmark                   & \cmark               \\
MySQL            & \cmark                   & \cmark               \\
\end{tabular}
\end{table}

Three ports are open, if the factory default settings are applied.
Table \ref{tab_defaultports} shows the results of a nmap scan of the \ac{PLC}s.
A vulnerability in one of the protocols leads to high risks, caused by the remote exploitability.

\begin{table}[H]
\centering
\caption{Factory Default Port Scan}
\label{tab_defaultports}
\begin{tabular}{c c c c}
\hline \hline
\textbf{Port} & \textbf{Protocol} & \textbf{State} & \textbf{Service} \\
\hline
21    & \acs{TCP}      & open  & FTP \\
1962  & \acs{TCP}      & open  & unknown \\
41100 & \acs{TCP}      & open  & unknown \\
\end{tabular}
\end{table}

\begin{figure}[H]
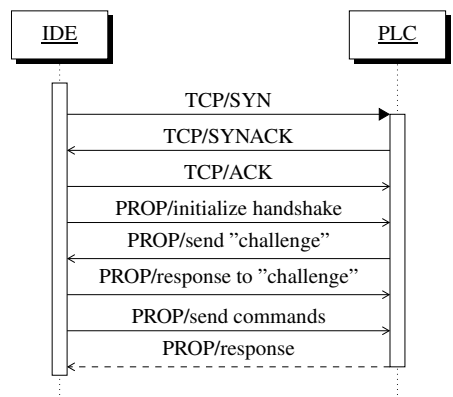

  \centering
  \scalebox{.8}[0.8]{
\begin{sequencediagram}
\newthread[white]{i}{IDE}
\newinst[4]{p}{PLC}

\begin{call}{i}{TCP/SYN}{p}{PROP/response}
\mess{p}{TCP/SYNACK}{i}
\mess{i}{TCP/ACK}{p}
\mess{i}{PROP/initialize handshake}{p}
\mess{p}{PROP/send "challenge"}{i}
\mess{i}{PROP/response to "challenge"}{p}
\mess{i}{PROP/send commands}{p}
\end{call}

\end{sequencediagram}
}
\caption{Handshake between Phoenix \ac{PLC} and \ac{IDE}}
\label{fig_handshake}
\end{figure}

For our purpose the most interesting ports are the undocumented ones, which are used by the \ac{IDE} to communicate with the \ac{PLC}.
Most commands are exchanged via port 1962.
The connection establishment of this protocol is illustrated in Figure \ref{fig_handshake}.
The \ac{IDE} sends a \ac{SYN} request to the \ac{PLC}.
This should respond with a \ac{SYN} \ac{ACK}.
Consequentially the \ac{IDE} complete the \ac{TCP} handshake with an \ac{ACK}.

After the TCP handshake with the \ac{PLC}, a proprietary initializing sequence between the \ac{IDE} and the Phoenix Contacts \ac{PLC} is necessary.
This starts with a request from the \ac{IDE} to the controller:
\begin{table}[H]
\centering
\texttt{
\begin{tabular}{K{0.8cm} K{5cm} K{1.2cm}}
0000   & 01 01 00 1a 00 00 00 80   & ........ \\
0008   & 64 15 00 03 00 0c 49 42   & d.....IB \\
0010   & 45 54 48 30 31 4e 30 5f   & ETH01N0\_ \\
0018   & 4d 00                     & M. \\
\end{tabular}
}
\end{table}

This request is answered from the \ac{PLC}, with an identifier (in this case 0x48) in it:

\begin{table}[H]
\centering
\texttt{
\begin{tabular}{K{0.8cm} K{5cm} K{1.2cm}}
0000   & 81 01 00 14 00 00 00 01   & ........ \\
0008   & 00 00 00 00 00 02 00 00   & ........ \\
0010   & 00 \textbf{48} 00 00               & .\textbf{H}.. \\
\end{tabular}
}
\end{table}

This value must be send back from the \ac{IDE} to the \ac{PLC}.
It could be seen as a simple session key, which does not change on a controller:

\begin{table}[H]
\centering
\texttt{
\begin{tabular}{K{0.8cm} K{5cm} K{1.2cm}}
0000   & 01 05 00 16 00 01 00 00   & ........ \\
0008   & e8 e9 00 \textbf{48} 00 00 00 1c   & ...\textbf{H}.... \\
0010   & 00 04 02 95 00 00         & ...... \\
\end{tabular}
}
\end{table}

For the proprietary protocol from Phoenix Contacts the same handshake for a specific \ac{DuT} could be send, because it is a constant value for each device.
Thus, the handshake is a simple replay for the Phoenix \ac{PLC}s and PropFuzz is able to detect the handshake by statistically comparing the start sequences of live captures or pcap-ng files.
After the proprietary handshake, commands could be send to the controller.
Below a reset command is shown, where the \ac{PLC} performs a complete reboot.

\begin{table}[H]
\centering
\texttt{
\begin{tabular}{K{0.8cm} K{5cm} K{1.2cm}}
0000  & 01 05 00 16 00 10 00 00  & ........ \\
0008  & e8 c8 00 \textbf{48} 00 00 00 00  & ...\textbf{H}.... \\
0010  & 00 04 0a ba 00 00        & ......   \\
\end{tabular}
}
\end{table}

At this point scapy is used to fuzz different fields of the command.
With the PropFuzz framework it was possible to identify different vulnerabilities in the session management and command handling of the \acp{PLC}.
The vulnerabilities found in the Phoenix Contact products were reported to the manufacturer and customers were informed with an advisory (ICSA-16-313-01).

\section{Possible Attacks \& Recommendations}
\label{sec:attacks}
The identified security problems within the protocol make several remote attacks possible.
Considering the usage of these controllers in critical infrastructures, the severity of potential attacks is classified  high.
\begin{itemize}
\item A \textbf{replay attack} is a network attack in which a valid data transmission is repeated.
The attacker needs few knowledge and can simple replay previous captures containing \ac{PLC} commands.
\item By \textbf{manipulating variables}, it is possible to change the sequence or process of a program on the \ac{PLC}.
This requires knowledge about the setup of the control system.
\item With \textbf{changing the software or firmware} of an \ac{ICS} the total control of it could be achieved, e.g. to create a BotNet.
\end{itemize}
These attacks demonstrate the severity of the found vulnerabilities with our framework.
It is possible to remotely exploit effected \ac{PLC}s without physical access to it.

\label{sec:recommendations}
The \textbf{recommendations} to avoid such vulnerabilities or to defend against possible attacks can be categorized by the responsible stake holders: manufacturer, system integrator and operator.
Integrators and end users of \ac{ICS}s should use a defense-in-depth architecture for their networks, considering the following:
\begin{itemize}
\item Devices should not be accessible public without the use of a \ac{VPN}
\item Firewalls should be used for network segmentation and controller isolation
\end{itemize}

Manufacturers of \ac{ICS}s should develop products, which are secure by design, e.g.:
\begin{itemize}
\item Authentication for the communication
\item Well implemented session and user management
\item Secure and cryptographically protected protocols
\end{itemize}

\section{Outlook \& Conclusion}
\label{sec:outlook}
The PropFuzz framework has reached a stable testing state.
We have already used our framework for fuzzing \ac{PLC}s from other vendors.
For similar protocols, PropFuzz is able to perform the handshake and start fuzzing without any adjustment to our framework.
In the current implementation, complex protocols with a proper session management and cryptographic measures are not fuzzable.
This functionality will be extended later, making it possible to fuzz a wider spectrum of protocols.
Furthermore, to observe a \ac{PLC} within a process it is necessary to virtually represent and observe all used in- and outputs of the controller.
This must be done with high efforts for every simulated process itself, which is not feasible at this time.

In this paper, we presented a stable and extensible fuzzing framework for proprietary \ac{ICS} protocols.
Compared with the available software, PropFuzz is able to automatically analyze the communication between the \ac{IDE} and the \ac{PLC} and fuzz the \ac{DuT}.
In addition, PropFuzz is able to monitor the output and detect suspicious behavior.

We have demonstrated the abilities of our framework by fuzzing two \ac{PLC}s
and detected three critical vulnerabilities, which could be exploited remotely by attackers (Advisory ICSA-16-313-01).
We have worked in a close cooperation with Phoenix Contacts to find solutions and fixes.

\section*{Project Funding}
The work on proprietary \ac{ICS} protocol fuzzing is part of the RiskViz \cite{Riskviz} research project.
This is funded by the Federal Ministry of Education and Research (BMBF), with the aim of creating a risk map of SCADA systems in Germany.

\bibliographystyle{ieeetr}
\bibliography{\jobname}

\begin{acronym}
 \acro{ACK}{acknowledgment}
 \acro{ARP}{Address Resolution Protocol}
 \acro{DuT}{Device under Test}
 \acro{ERP}{Enterprise Resource Planning}
 \acro{ICS}{Industrial Control System}
 \acrodefplural{ICS}{Industrial Control Systems}
 \acro{IDE}{Integrated Development Environment}
 \acro{IoT}{Internet of Things}
 \acro{IP}{Internet Protocol}
 \acro{MES}{Manufacturing Execution System}
 \acro{PLC}{Programmable Logic Controller}
 \acrodefplural{PLC}{Programmable Logic Controllers}
 \acro{SCADA}{Supervisory Control and Data Acquisition}
 \acro{SYN}{synchronize}
 \acro{TCP}{Transmission Control Protocol}
 \acro{USB}{Universal Serial Bus}
 \acro{VPN}{Virtual Private Network}
\end{acronym}

\end{document}